\documentclass[letterpaper, 10 pt, conference]{ieeeconf}  
\IEEEoverridecommandlockouts                             
\usepackage[english]{babel}
\usepackage{amsmath, amssymb, amsfonts, bm}
\usepackage{mathrsfs}
\usepackage{textcomp}
\usepackage[mathcal]{euscript}
\usepackage{enumerate}
\usepackage{mathbbol}
\usepackage{float}
\usepackage{color,graphicx,epstopdf}
\usepackage{dsfont}
\usepackage{hyperref}
\usepackage[noadjust]{cite}
\usepackage{caption}
\usepackage{subcaption}
\usepackage{mathtools}
\usepackage{tikz-cd}
\usepackage{mathtools}
\usepackage{algorithm}
\usepackage{algpseudocode}
\usepackage{url}
\usepackage{authblk}
\usepackage{verbatim}
\usepackage{stackengine}
\usepackage{subcaption}
\usepackage{xcolor}

\newtheorem{theorem}{Theorem}
\newtheorem{definition}{Definition}

\newtheorem{lemma}{Lemma}
\newtheorem{remark}{Remark}
\newtheorem{proposition}{Proposition}
\newtheorem{assumption}{Assumption}

\usepackage[left=1.91cm,right=1.91cm,top=1.91cm,bottom=1.91cm]{geometry}

\definecolor{green}{rgb}{0.05, 0.5, 0.06}

\newcommand{\mcalV}{\mathcal{V}}
\newcommand{\mcalN}{\mathcal{N}}
\newcommand{\mcalE}{\mathcal{E}}

\newcommand{\bx}{\mathbf{x}}
\newcommand{\bA}{\mathbf{A}}
\newcommand{\bB}{\mathbf{B}}
\newcommand{\bC}{\mathbf{C}}

\newcommand{\bQ}{\mathbf{Q}}
\newcommand{\bG}{\mathbf{G}}
\newcommand{\by}{\mathbf{y}}
\newcommand{\bu}{\mathbf{u}}
\newcommand{\bj}{\mathbf{j}}

\DeclareMathOperator{\diag}{diag}
\newcommand{\alldiag}{\stackrel{n}{\underset{i=1}{\diag}}}

\renewcommand{\color}[1]{}

\title{\LARGE \bf
	Design and Stability of {\color{red}Angle based} Feedback Control \\in Power Systems:
	{\color{red}A} Negative-Imaginary Approach
}

\author{Yijun Chen$^{1}$, Ian R. Petersen$^{1}$, and Elizabeth L. Ratnam$^{1}$
	
	\thanks{$^{1}$The  School of Engineering, The Australian National University, Canberra, Australia, emails: \{yijun.chen, ian.petersen, elizabeth.ratnam \}@anu.edu.au.}%
	
	\thanks{This work was supported by the Australian Research Council under grants DP230102443 and LP210200473.}
}

\begin{document}

\maketitle
\thispagestyle{empty}
\pagestyle{empty}

\begin{abstract}
	This paper considers a power transmission network characterized by interconnected nonlinear swing dynamics on generator buses. {\color{red}At} the steady state, frequencies across different buses synchronize to a common nominal value {\color{red}such as $50$Hz or $60$Hz}, and power flows on transmission lines are within  {\color{red}steady-state} envelopes. We {\color{red}assume that fast measurements of generator rotor angles are available}. Our approach {\color{red}to frequency and angle control} centers on equipping generator buses with {\color{red}large-scale batteries that are controllable on a fast timescale}. We link {\color{red}angle} based feedback {\color{red}linearization} control with negative-imaginary systems theory. Angle {\color{red}based feedback controllers are} designed using {\color{red}large-scale} batteries as actuators and can be implemented in a distributed manner incorporating local information. Our analysis demonstrates the internal stability of the interconnection between the power transmission network and the {\color{red}angle based feedback controllers}. This internal stability underscores the benefits of achieving frequency synchronization and preserving {\color{red}steady-state} power flows within network envelopes {\color{red}through the use of feedback controllers}. {\color{red}Our approach will enable transmission lines to be operated at maximum power capacity since stability robustness is ensured by the use of feedback controllers rather than conservative criteria such as the equal area criterion.} By means of numerical simulations we {\color{red}illustrate} our {\color{red}results}.
\end{abstract}

\section{Introduction}
{\color{green} In the transition to net zero power systems, it has been suggested that a massive expansion of the transmission grid will be required to support emerging renewable energy zones \cite{AEMO_TE}. In this paper, we propose the use of feedback control and negative imaginary systems theory \cite{petersen2010feedback,lanzon2008stability,xiong2010negative,petersen2016negative,wang2015robust} to reduce the need for such an expansion by enabling the more complete utilization of existing grid infrastructure.}

Frequency deviations {\color{red}in electric power systems} result from imbalances between electrical load and generator output, serving as a key indicator of generation-load mismatches. Frequency deviations can harm electrical equipment or degrade performance, and can overload transmission lines thereby triggering protective measures --- all of which impacts power system operation and reliability \cite{bevrani2014robust}. Accordingly, frequency control aims to keep the frequency of power systems closely aligned with the designated standard {\color{red}of $50$Hz or $60$Hz} in the presence of imbalances between electrical demand and generator supply. 

The objective of frequency regulation can be realized from both the generation and load sides. Automatic generation control, which adjusts generator setpoints in response to variations in frequency and unexpected power {\color{red}flows} between different areas, is a traditional {\color{red}frequency regulation} scheme from the generation side \cite{ilic2012toward,wood2013power}. Alternatively, with the deepening integration of renewable energy {\color{red}sources}, increased volatility in non-dispatchable renewable generation has imposed challenges on generator-side control. To address these challenges, load control has received considerable attention in  recent years \cite{liu2013decentralized,beil2016frequency,kim2014modeling}.

Despite the successful establishment of frequency control techniques in maintaining the nominal frequency, it falls short in achieving precise power flow control on transmission lines. Moreover, existing frequency control techniques prove insufficient in maintaining stability as renewable energy zones continue to displace fossil-fuel generators and thereby reduce the overall system inertia in the bulk power grid. Furthermore, the indiscriminate adjustment of power flow can inadvertently push transmission lines perilously close to their {\color{red}transient stability} limits, posing a significant risk to the integrity of the power {\color{red}system} infrastructure \cite{gonen2009electrical,doukas2011electric}. The long-term consequences can manifest as physical damage to vital components within the power system {\color{red}and even the initiation of cascading failure events}.

In recent years, advancements in battery technologies have led to the widespread adoption of {\color{red}rechargeable batteries} in electric vehicles, {\color{red}the use of large grid storage batteries}, and the proliferation of domestic solar-powered batteries \cite{tran2019efficient,borenstein2022s}. {\color{red}In addition to} energy storage supporting local power management, this transformative development may also usher in the possibility of leveraging energy storage systems for active participation in both frequency and power flow regulation {\color{red}as well as guaranteeing transient stability} within power systems. However, the integration of distributed energy storage {\color{red}devices} into vital {\color{red}power system} control mechanisms requires a deep understanding of the interaction {\color{red}between rapidly controlled batteries and} the dynamic behavior of power transmission networks, paving the way for ubiquitous continuous fast-acting distributed energy storage participation in both frequency and power flow regulation {\color{red}along with transient stability augmentation}. In pursuit of this objective, this paper provides a systematic method to design {\color{red}power system} feedback controllers {\color{red}based on the use of battery-based actuators} that help power transmission networks synchronize bus frequencies and preserve {\color{red}steady-state} power flows {\color{red}along with ensuring the transient stability of the system}.

The {\color{red}availability} of measured variables affects how controllers are designed for {\color{red}power systems}. In the literature {\color{red}on} generator-side frequency control, droop control is a conventional method, which senses {\color{red}the rotor speed of generators}, adjusts the input value from speed data to modulate the mechanical power output, and effectively restores bus frequencies to their designated nominal values \cite{bevrani2014robust,liu2015comparison}. In contrast to the conventional rotor speed feedback method, an alternative approach {\color{red}is} rotor angle feedback, {\color{red}which has been} previously limited by angle sensor technology. However, the evolution of rotor angle measurement {\color{red}methods} holds {\color{red}the possibility of} real time angle {\color{red}measurements for generators} \cite{vivsic2020synchronous}. Importantly, the development {\color{red}of fast} rotor angle {\color{red}measurements}  presents a {\color{red}significant} opportunity for future power systems, {\color{red}enabling the control of frequency and power flows, and ensuring the transient stability through the use of feedback control. This will, in turn, enable power systems to be utilized at their maximum capacity without the need for over-engineering to ensure transient stability.} Negative-imaginary (NI) systems theory \cite{petersen2010feedback,lanzon2008stability,xiong2010negative,petersen2016negative,wang2015robust} can be utilized to guarantee the internal stability of interconnected power systems, ensure system robustness, and address consensus problems relating to the convergence of bus angles.

In this paper, we investigate the application of {\color{red}angle based} feedback {\color{red}linearization control to power transmission networks using NI systems theory to guarantee power system stability and synchronization.} We consider a power transmission network  characterized by interconnected nonlinear swing dynamics on generator buses. At steady state, the frequencies on different buses are synchronized to a standard value, and power flows on transmission lines are operated at {\color{red}steady-state} levels. We make the assumption that the rotor angles can be measured in real time. Rather than the participation of controllable load in frequency regulation, our focus lies in equipping generator buses with {\color{red}large-scale} batteries that {\color{red}can act as actuators for our angle based controllers}.

Our contributions are as follows. We design {\color{red}angle based feedback linearization controllers using real time angle sensors and large-scale batteries as actuators}. In particular, the proposed controllers provide three advantages: 1) based on NI systems theory, the interconnection of the power transmission network and the {\color{red}angle based} feedback controllers is proved to be internally stable; 2) the frequencies on different generator buses are synchronized to a nominal value and the power flows on transmission lines are maintained at {\color{red}steady-state} levels before a fault; 3) with local measurement and local communication of rotor angle {\color{red}measurements}, {\color{red}the overall control system} operates in a fully distributed manner. It is noted that although we refer to the feedback controllers as `angle based feedback controllers', the actual accessible measurement data is assumed to be rotor angle data (see Remark~\ref{rmk:local_storage}).

This paper is organized as follows. Section~\ref{sec:preliminary} provides {\color{red}some} preliminary knowledge on negative-imaginary systems. Section~\ref{sec:transmission_network} elaborates on a nonlinear dynamic model for power transmission networks. Section~\ref{sec:design} applies negative-imaginary systems theory to design controllers based on angle feedback. Section~\ref{sec:examples} gives simulation results, {\color{red}which illustrate our theory}. Section~\ref{sec:conclusion} concludes the paper.

\section{Preliminary results on Negative-Imaginary Systems}\label{sec:preliminary}
In this section, we present {\color{red}some} preliminary {\color{red}material on} negative-imaginary systems.

\medskip

\begin{definition}[\cite{petersen2010feedback}]\label{def:NI}
	The square transfer function matrix $M(s)$ is negative-imaginary (NI) if the following conditions are satisfied:
	\begin{itemize}
		\item [(1)] all of the poles of $M(s)$ lie in the open left half of the complex plane (OLHP),
		
		\item [(2)] for all $\alpha \geq 0$,
		\begin{equation} \label{eq:NI_con2}
			\bj[M(\bj \alpha) - M^{\ast}(\bj \alpha)] \geq0.
		\end{equation}
	\end{itemize}
	A linear time-invariant system is NI if its transfer function matrix is NI.
\end{definition}

\medskip

\begin{definition}[\cite{petersen2010feedback}]\label{def:SNI}
	The square transfer function matrix $M(s)$ is strictly negative-imaginary (SNI) if the following conditions are satisfied:
	\begin{itemize}
		\item [(1)] all of the poles of $M(s)$ lie in the OLHP,
		
		\item [(2)] for all $\alpha > 0$,
		\begin{equation} \label{eq:SNI_con2}
			\bj[M(\bj \alpha) - M^{\ast}(\bj \alpha)] >0.
		\end{equation}
	\end{itemize}
	A linear time-invariant system is SNI if its transfer function matrix is SNI.
\end{definition}

\medskip

\begin{lemma}[\cite{petersen2010feedback}]\label{lemma:SNI}
	Consider the NI transfer function matrix $M(s)$ and the SNI transfer function matrix $N(s)$, and suppose that $M(\infty)N(\infty) = 0$ and $N(\infty) \geq 0$. Then, the positive-feedback interconnection of $M(s)$ and $N(s)$ is internally stable if and only if 
	\begin{equation}\label{eq:stable_condition}
		\lambda_{max}(M(0)N(0)) < 1.
	\end{equation}
\end{lemma}

\medskip

\begin{lemma}[\cite{lanzon2008stability}]\label{lemma:zero_frequency}
	Given an NI (resp. SNI) transfer function {\color{red}$M(s)$,} then $M(0) - M(\infty) \geq 0$ (resp. $>0$).
\end{lemma}

\medskip

\begin{lemma}[\cite{wang2015robust}]\label{lemma:lambda_inequality}
	Given a matrix $M$ which is Hermitian with $\lambda_{\max}(M) \geq 0$ and a Hermitian matrix $N \geq 0$, we have $\lambda_{\max}(MN) \leq \lambda_{\max}(M) \lambda_{\max}(N)$.
\end{lemma}

\section{Transmission Network Model}\label{sec:transmission_network}
Consider a transmission network comprised of $n$ generator buses and $l$ transmission lines. The  topology of the power transmission network is described by a connected and undirected graph $\mathcal{G} = (\mathcal{V},\mathcal{E})$, where $\mathcal{V} = \{1,2,\dots,n\}$ is the set of buses, and $\mathcal{E} \subseteq \mathcal{V} \times \mathcal{V}$ is the set of transmission lines connecting the buses. If there is a transmission line connecting bus $i \in \mathcal{V}$ and bus $j \in \mathcal{V}$, then we say bus $i$ and bus $j$ are neighbors with $(i,j) \in \mathcal{E}$. The neighboring buses of bus $i \in \mathcal{V}$ are indexed in the set $\mathcal{N}(i) = \{j | (i,j) \in \mathcal{E}\}$. The adjacency matrix of the transmission network is denoted by $\mathbf{T} \in \mathbb{R}^{n \times n}$, where $T_{ij} = 1$ if $(i,j) \in \mathcal{E}$, and $T_{ij} = 0$ if $(i,j) \notin \mathcal{E}$. We choose a fixed representation of transmission lines. Each $(i,j)$ or $(j,i)$ can only be chosen once. Slightly different from the graph literature, the incidence matrix $\mathbf{Q} \in \mathbb{R}^{n \times l}$ of the transmission network is then defined by 
\begin{equation*}
	\mathbf{Q} := 
	\begin{cases}
		q_{ie} = 1, & \text{if } i \text{ is the initial node of edge } e,\\
		q_{ie} = -1, & \text{if } i \text{ is the terminal node of edge } e,\\
		q_{ie} = 0, & \text{if } i \text{ is not connected in edge } e.
	\end{cases}
\end{equation*}
Note that the slight modification of the incidence matrix is designed for the ease of calculation in later sections, and ``initial/terminal node'' does not refer to a particular orientation.

\begin{assumption}\label{apt:power_network}
	We adopt the following assumptions that are well-justified for power transmission networks in \cite{kundur2022power, zhao2012swing}.
	\begin{enumerate}
		\item Each transmission line $(i,j) \in \mathcal{E}$ is lossless and only characterized by its reactance $ X_{ij}$.
		
		\item The internal voltage magnitudes $E_{i}^{0}$ of each bus $i \in \mathcal{V}$ are constant.
		
		\item {\color{blue}Reactive power injections on the buses are controlled to maintain $E_{i}^{0}$ of each generator bus $i \in \mathcal{V}$, and such controllers are decoupled from frequency regulation in transmission grids \cite{rathnayake2022multivariable}.}
	\end{enumerate}
	However, note that we do not assume that {\color{red}the rotor angle difference between generators is small.}
\end{assumption}

\subsection{Network Model} 
{\color{red}In what follows, we consider only generator buses.} {\color{green}However, in practice, some buses could be distributed load buses without a physical generator. In this case, the same dynamics could be obtained by using grid-forming inverters \cite{lasseter2019grid} with distributed energy resources such as batteries and solar generation. The corresponding `rotor' angle would be determined from the inverter voltage waveform as in~\eqref{eq:rotor_phase_relationship} below.} In this case, simulated inertia could be achieved if the power delivered by the grid-forming inverter is proportional to the negative of the rate of the change of frequency (ROCOF). 
\medskip

\begin{figure}[tb]
	\centering
	\includegraphics[width=0.35\textwidth]{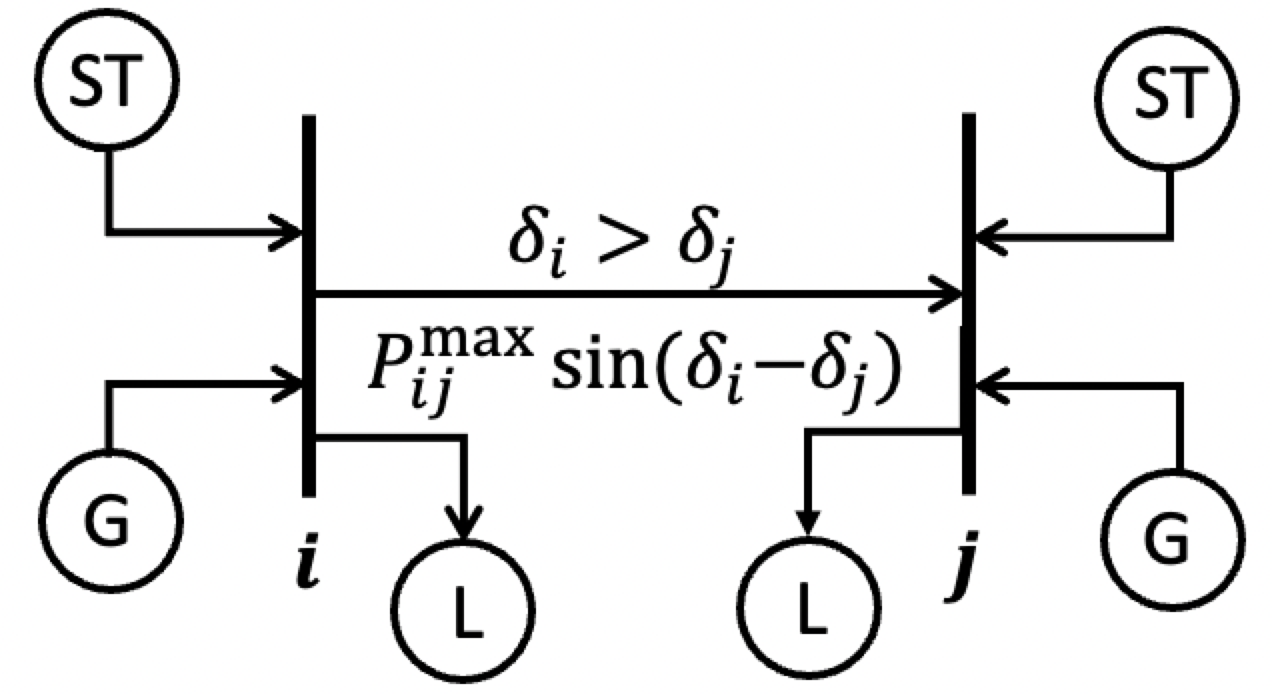}
	\caption{The components of a generator bus where `G' represents an AC generator that converts mechanical power into electric power, `L' represents fixed inflexible load that consumes electric power, and `ST' represents a local storage device that outputs {\color{red}stored excess power}.}
	\label{fig:components}
\end{figure}

As depicted in Fig.~\ref{fig:components}, a generator bus has an AC generator that converts mechanical power into electric power through a rotating prime mover, fixed inflexible load that consumes {\color{red}a known amount of} electric power, and a local storage device that outputs storage power. For each generator bus $i \in \mathcal{V}$, we denote the rotor angle in the stationary reference frame by $\delta_{i}$, the rotor speed by $\omega_{i}$, and the nominal frequency by $\omega^{0}$. Then, each generator bus $i$'s voltage at time $t$ is written as a cosine function multiplied by its  voltage magnitude $|V_{i}|$:
\begin{subequations}\label{eq:rotor_phase_relationship}
	\begin{align}
	|V_{i}|\cos(\delta_{i}(t)) &= |V_{i}|\cos(\omega^{0}t + \theta(t)) \\
	&= |V_{i}|\cos(\omega^{0}t + \bar{\theta}_{i} + \tilde{\theta}_{i}(t)),
	\end{align}
\end{subequations}
where $\theta(t)$ represents the time-varying phase angle, $\bar{\theta}_{i}$ represents the phase angle {\color{red}at steady state}, and $\tilde{\theta}_{i}(t)$ represents the time-varying phase angle deviation. The frequency on each generator bus $i$ is defined as 
\begin{equation}\label{eq:frequency_def}
	\omega_{i} = \omega^{0} + \dot{\tilde{\theta}}_{i}.
\end{equation}

For each generator bus $i \in \mathcal{V}$, its dynamics are modeled by the following swing equations \cite{chen2020distributed, dorfler2012synchronization}
\begin{subequations}\label{eq:swing_dynamics}
	\begin{align}
		\dot{\delta_{i}} &= \omega_{i}\\
		\dot{\omega_{i}} &= \frac{\omega^{0}}{2H_{i}}\Big[P^{ M}_{i} + P^{ ST}_{i} - \frac{D'_{i}}{\omega^{0}}\omega_{i}  - P^{ E}_{i}\Big],
	\end{align}
\end{subequations}
where $H_{i} > 0$ is the inertia coefficient of generator bus $i$, and $D'_{i} > 0$ is the damping power coefficient of generator bus $i$. Here, $P^{ M}_{i}$ represents the mechanical power injection to generator bus $i$, $P^{ ST}_{i}$ represents the power output of the local storage device at generator bus $i$, and $P^{ E}_{i}$ represents the electric power output of generator bus $i$. The electric power output of generator bus $i$ equals the sum of the inflexible power consumption of the load at generator bus $i$ and the net branch power flow from generator bus $i$ to other neighboring buses, which is described by 
\begin{equation}\label{eq:electric_output}
	P^{ E}_{i} = P^{ L}_{i}  + \sum_{j \in \mathcal{N}(i)}P_{ij},
\end{equation}
where $P^{ L}_{i}$ is the known fixed inflexible load power, and $P_{ij}, (i,j) \in \mathcal{E}$ is the power flow from generator bus $i$ to a neighboring bus $j \in \mathcal{N}(i)$. With Assumption~\ref{apt:power_network}, the branch power flow on the branch $(i,j) \in \mathcal{E}$ is described by \cite[Chapter 7.10]{machowski1997power}, \cite{rathnayake2022multivariable}
\begin{equation}\label{eq:branch_flow}
	P_{ij} =  \frac{E_{i}^{0}E_{j}^{0}}{X_{ij}}\sin(\delta_{i} - \delta_{j}) = P_{ij}^{\max}\sin(\delta_{i} - \delta_{j}), 
\end{equation}
where $X_{ij}$ is the reactance of the transmission line $(i,j)$, and $P_{ij}^{\max} = \frac{E_{i}^{0}E_{j}^{0}}{X_{ij}}$ is the maximum power transfer on the transmission line $(i,j)$.

Combining Eqs.~\eqref{eq:swing_dynamics}, ~\eqref{eq:electric_output}, and~\eqref{eq:branch_flow}, the interconnected swing equations for the transmission network in terms of rotor angles $\delta_{i}, \forall i \in \mathcal{V}$ are then given by 
\begin{equation}\label{eq:delta_interconnected_swing}
	M_{i}\ddot{\delta_{i}} + D_{i}\dot{\delta_{i}} = P^{ M}_{i} + P^{ ST}_{i}   - P^{ L}_{i} - \sum_{j \in \mathcal{N}(i)}P_{ij}^{\max}\sin(\delta_{i} - \delta_{j}), 
\end{equation}
with $M_{i} = \frac{2H_{i}}{\omega^{0}} > 0$ and $D_{i} = \frac{D'_{i}}{\omega^{0}} > 0$ for all $i \in \mathcal{V}$.

\medskip

\begin{remark}
	With increasing {\color{red}adoption of} renewable energy resources, the availability of supporting battery technology has become {\color{red}widespread} such as in mobile batteries in electric vehicles and in {\color{red}domestic storage} batteries. {\color{red}Also,} in the future, it is expected that large-scale battery systems will be deployed {\color{red}at} buses to store surplus power {\color{red}from wind and solar energy resources}. {\color{red}In addition to} storage purposes, this paper {\color{red}proposes the use of such large-scale batteries as actuators in power system controllers} to 
	stabilize {\color{red}the power} system, synchronize bus frequencies, and maintain steady-state power flows when a fault happens.
\end{remark}
\medskip

Before the occurrence of a fault {\color{red}in a power system}, the stable equilibrium  of the system \eqref{eq:delta_interconnected_swing} is denoted by
\begin{equation}\label{eq:stable_equilibrium}
(\bar{\dot{\delta}}_{i}, \bar{P}_{i}^{M}, \bar{P}_{i}^{L}, \bar{P}_{i}^{ST}, \bar{P}_{ij}^{\max}, \bar{\psi}_{ij}),  \forall i \in \mathcal{V},  \forall (i,j) \in \mathcal{E},
\end{equation}
with
\begin{equation}\label{eq:def_nominal_diff}
	\bar{\psi}_{ij} = \bar{\theta}_{i} - \bar{\theta}_{j}
\end{equation}
representing the {\color{red}steady-state} rotor angle difference between two neighboring buses $i$ and $j$. The relation between these steady-state values is described by 
\begin{equation}\label{eq:equilibrium}
	D_{i}\omega^{0} = D_{i}\bar{\dot{\delta}}_{i}  = \bar{P}^{ M}_{i} +\bar{ P}^{ ST}_{i}   - \bar{P}^{ L}_{i} - \sum_{j \in \mathcal{N}(i)}\bar{P}_{ij}^{\max}\sin\bar{\psi}_{ij}.
\end{equation}
{\color{red}In the aftermath of a fault} in the transmission network, the bus frequencies $\omega_{i},\forall \in \mcalV$ and angle differences $\delta_{i} - \delta_{j}, \forall (i,j) \in \mcalE$ deviate away from their steady-state values. {\color{red}We seek to restore the bus frequencies and angle differences to their steady-state values}. In {\color{red}what follows}, we investigate the post-fault transients, during which mechanical power, load power consumption, and maximum power transfer are all back to their nominal values in Eq.~\eqref{eq:stable_equilibrium}. We rewrite interconnected swing equations in terms of phase angle deviation $\tilde{\theta}_{i}, \forall i \in \mathcal{V}$ into state-space form, and design {\color{red}angle based} feedback controllers for local storage devices at generator buses to stabilize the power transmission network, regulate the frequencies on generator buses, and maintain {\color{red}steady-state} power flow on transmission lines.

\subsection{State-Space Model} Here, we present the reformulation of the network model into the state-space form.
\medskip

For each generator bus $i \in \mcalV$, we define $\tilde{P}_{i}^{ST} = P_{i}^{ST} - \bar{P}_{i}^{ST}$ as the storage power output deviation from the  storage power output {\color{red}at steady state}. By using Eqs.~\eqref{eq:rotor_phase_relationship}, \eqref{eq:frequency_def}, \eqref{eq:delta_interconnected_swing}, and \eqref{eq:equilibrium}, the interconnected swing equations can be rewritten in terms of the phase angle deviation $\tilde{\theta}_{i}$ specifically:
\begin{equation}\label{eq:swing_deviation}
	\begin{aligned}
		M_{i}\ddot{\tilde{\theta}}_{i} + D_{i}\dot{\tilde{\theta}}_{i} + K_{i}\tilde{\theta} &=  K_{i}\tilde{\theta}_{i} + \tilde{P}^{ ST}_{i} +  \sum_{j \in \mcalN(i)}\bar{P}_{ij}^{\max}\sin\bar{\psi}_{ij} \\
		& - \sum_{j \in \mcalN(i)}\bar{P}_{ij}^{\max}\sin(\tilde{\theta}_{i} - \tilde{\theta}_{j} + \bar{\psi}_{ij}) , 
	\end{aligned}
\end{equation}
where $K_{i} > 0$ is arbitrarily chosen as a positive real number. {\color{red}Although} $K_{i}$ does not change the nature of the power transmission system, it can help construct an SNI plant (see in Section \ref{sec:SNI_plant}).  

\medskip

{\bf \noindent Generator Buses.} For each generator bus $i \in \mcalV$, we define the state as $\bx_{i} =  [\dot{\tilde{\theta}}_{i}, \tilde{\theta}_{i}]^{\top}$ and the output as $\by_{i} = [\tilde{\theta}_{i}] $. According to each generator bus system~\eqref{eq:swing_deviation}, the state-space model for each generator bus $i \in \mcalV$  is given by 
\begin{subequations}\label{eq:single_sys}
	\begin{align}
		\dot{\bx}_{i}& = \bA_{i}\bx_{i} + \bB_{i}\bu_{i},\\
		\by_{i}& = \bC_{i}\bx_{i},
	\end{align}
\end{subequations}
with system matrices 
\begin{align*}
	\bA_{i} &= 	\begin{bmatrix}
		-\frac{D_{i}}{M_{i}} & -\frac{K_{i}}{M_{i}}\\
		1 & 0
	\end{bmatrix},\\
	\bB_{i} &=  \begin{bmatrix}
		\frac{1}{M_{i}}\\
		0
	\end{bmatrix},\\
	\bC_{i} &= \begin{bmatrix}
		0 & 1
	\end{bmatrix},
\end{align*}
and the input 
\begin{equation}\label{eq:input}
	\bu_{i} = K_{i}\tilde{\theta}_{i} + \tilde{P}^{ ST}_{i} +  \sum_{j \in \mcalN(i)}\bar{P}_{ij}^{\max}\big(\sin\bar{\psi}_{ij} -\sin(\tilde{\theta}_{i} - \tilde{\theta}_{j} + \bar{\psi}_{ij})\big).
\end{equation}
For each generator bus $i \in \mcalV$, the transfer function $G_{i}(s)$ from the input $\bu_{i}$ to the output $\by_{i}$ is described by 
\begin{equation}\label{eq:tf}
	G_{i}(s) = \frac{1}{M_{i}s^{2} + D_{i}s + K_{i}}.
\end{equation}

In practice, any input $\bu_{i}$ can be realized by setting the output power deviation of the local storage device at generator bus $i$ as 
\begin{equation}\label{eq:local_storage}
	\tilde{P}^{ ST}_{i} = 
	\bu_{i}  -  \sum_{j \in \mcalN(i)}\bar{P}_{ij}^{\max}\big(\sin\bar{\psi}_{ij} -\sin(\tilde{\theta}_{i} - \tilde{\theta}_{j} + \bar{\psi}_{ij})\big) - K_{i}\tilde{\theta}_{i}.
\end{equation}

\begin{remark}\label{rmk:local_storage}
	The {\color{red}use} of Eq.~\eqref{eq:local_storage} for dispatching power from the local storage device enables us to cancels out the non-linearity in the power transmission network by feedback linearization. With measurements of the rotor angle $\delta_{i}$ and the steady-state phase angle $\bar{\theta}_{i}$, we compute $\tilde{\theta}_{i} = \delta_{i} - \omega^{0}t - \bar{\theta}_{i}$, which is applied as feedback in Eq.~\eqref{eq:local_storage} . However, it is also {\color{red}important} to note that {\color{red}this will require fast rotor angle  measurements ($10^{-4}$ seconds) \cite{vivsic2020synchronous}. Also, in practice, the battery storage output will be subjected to actuator saturation. {\color{green} This will mean that in practice only a finite margin of transient stability can be achieved depending on the power of the inverter actuator. However, even if the transmission lines are operated at maximum power, a good level of transient stability margin can be achieved by a suitable choice of control actuator inverter. This is in contrast to the situation in the current grid in which the utilization of transmission lines is limited via the requirement for a transient stability margin as determined by the equal area criterion (see \cite{machowski1997power}).}}
\end{remark}
\medskip

\begin{figure}[tb]
	\centering
	\includegraphics[width=0.32 \textwidth]{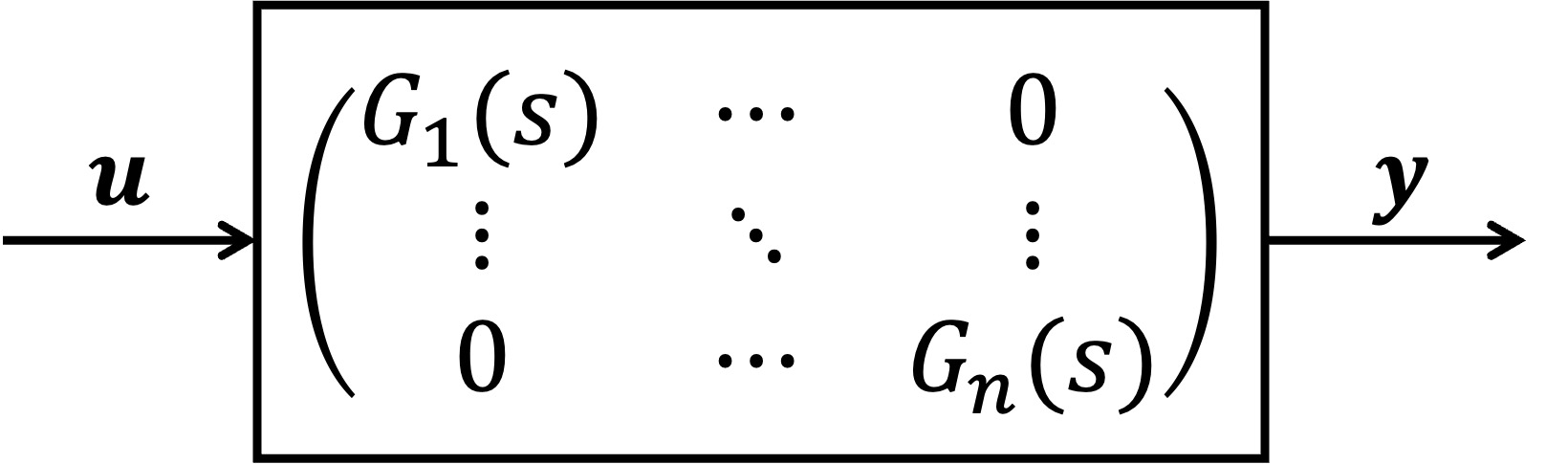}
	\caption{The overall plant for the transmission network.}
	\label{fig:overall_plant}
\end{figure}

{\bf \noindent Transmission Network.}  Define $\bx = [\bx_{1}^{\top}, \dots, \bx_{n}^{\top}]^{\top} \in \mathbb{R}^{2n}$,  $\by = [\by_{1}^{\top}, \dots, \by_{n}^{\top}]^{\top} \in \mathbb{R}^{n}$, and $\bu = [\bu_{1}^{\top}, \dots, \bu_{n}^{\top}]^{\top} \in \mathbb{R}^{n}$. The state-space model for the overall transmission network is  described by 
\begin{subequations}\label{eq:overall_sys}
	\begin{align}
		\dot{\bx}& = \bA\bx + \bB\bu,\\
		\by& = \bC\bx
	\end{align}
\end{subequations}
with system matrices 
\begin{align*}
	\bA = \alldiag\{\bA_{i}\},
	\bB = \alldiag\{\bB_{i}\},
	\bC = \alldiag\{\bC_{i}\}.
\end{align*}
The overall plant \eqref{eq:overall_sys} for the transmission network is illustrated in Fig.~\ref{fig:overall_plant}. The transfer function $\bG(s)$ from the input $\bu$ to the output $\by$ is given by
\begin{equation}\label{eq:overall_tf}
	\bG(s) = \alldiag\{G_{i}(s)\}.
\end{equation}

\subsection{SNI Plant}\label{sec:SNI_plant}
In what follows, we prove that each generator bus plant~\eqref{eq:single_sys} is SNI, and the overall plant~\eqref{eq:overall_sys} for the transmission network is SNI.

\begin{lemma}\label{lemma:sub_SNI}
	For each generator $i \in \mcalV$, the system~\eqref{eq:single_sys} is an SNI system.
\end{lemma}
{\it Proof.} Firstly, since $M_{i} > 0$, $D_{i} > 0$, and $K_{i} > 0$, we know both of the two poles of the transfer function $G_{i}(s)$ have negative real parts. The condition (1) in Definition \ref{def:SNI} is thus satisfied. Secondly, we can obtain for all $\alpha > 0$, 
\begin{equation*}
	\bj[G_{i}(\bj \alpha) - G_{i}^{\ast}(\bj \alpha)] = \frac{2D_{i}\alpha}{(-M_{i}\alpha^{2} + K_{i})^{2} + (D_{i}\alpha)^{2}} > 0.
\end{equation*}
The condition (2) in Definition \ref{def:SNI} is thus satisfied.
The proof is now completed. \hfill$\square$

\medskip

\begin{lemma}\label{lemma:overall_SNI}
	For the transmission network, the overall system \eqref{eq:overall_sys} comprised of multiple SNI systems \eqref{eq:single_sys} is an SNI system.
\end{lemma}
{\it Proof.} It is {\color{red}clear} that the poles of $\bG(s)$ lie in the OLHP. It is also true that for all $\alpha> 0 $, $$\bj[\bG(\bj \alpha) - \bG^{\ast}(\bj \alpha)] = \alldiag\{\bj[G_{i}(\bj \alpha) - G_{i}^{\ast}(\bj \alpha)]\} >0 .$$ The proof is now completed. \hfill$\square$

\section{ {\color{red}Angle based} Feedback Controllers}\label{sec:design}
In transmission networks, the occurrence of a fault may {\color{red}lead to transients that impact the operational stability, driving bus frequencies and power flows {\color{red}away from steady-state values} \cite{machowski1997power}.  In this section, we aim to design {\color{red}angle based} feedback controllers that are capable of maintaining system stability, synchronizing bus frequencies, and preserving the {\color{red}steady-state}  angle differences on transmission lines when the transmission network deviates from its stable equilibrium due to the occurrence of a fault.

\subsection{{\color{red}Angle based} Feedback Consensus}
We want the angle deviations $\by_{i} = \tilde{\theta}_{i} = \theta_{i} - \bar{\theta}_{i}, i \in \mcalV,$ to {\color{red}all converge to} a common value $y_{ss}$. Since
\begin{subequations}\label{eq:consensus}
	\begin{align}
		\by_{i} - y_{ss} &= \tilde{\theta}_{i} - y_{ss} = \theta_{i} - \bar{\theta}_{i} - y_{ss} \to 0, \label{eq:sub1}\\
		\by_{j} - y_{ss} &= \tilde{\theta}_{j} - y_{ss} = \theta_{j} - \bar{\theta}_{j} - y_{ss} \to 0, \label{eq:sub2}
	\end{align}
\end{subequations}
we obtain
\begin{equation}\label{eq:nominal_diff}
	(\theta_{i} - \theta_{j}) - (\bar{\theta}_{i} - \bar{\theta}_{j}) \to 0,
\end{equation}
by subtracting~\eqref{eq:sub1} and \eqref{eq:sub2}. Therefore, designing feedback controllers that {\color{red}control} $\by_{i}$ to {\color{red}converge to} a common value $y_{ss}$ for all $i \in \mcalV$ has the advantage of retaining angle differences on transmission lines as steady-state values.

\medskip

{\bf \noindent {\color{red}Angle based} feedback controllers.} In what follows, inspired by \cite{wang2015robust}, we {\color{red}consider} the angle based feedback system as depicted in Fig.~\ref{fig:overall_fb_plant}, where the output $\by$ reaches consensus.

We begin by defining a new output $\tilde{\by} = [\tilde{\theta}_{i} - \tilde{\theta}_{j}]_{(i,j) \in \mcalE} \in \mathbb{R}^{l}$, which can be easily obtained from {\color{red}the relation}:
\begin{equation}\label{eq:ytilde}
	\tilde{\by} = \bQ^{\top} \by.
\end{equation}
Note that when the  output $\tilde{\by} \in \mathbb{R}^{l} \to \mathbf{0}_{l}$, the real output $\by \in \mathbb{R}^{n}$ reaches consensus, i.e., $\by_{i} = \by_{j} = y_{ss}$ due to the fact that span$\{\mathbf{1}_{n}\}$ is in the null space of $\bQ^{\top}$.

Then, a new input $\tilde{\bu} \in \mathbb{R}^{l}$ is {\color{red}constituted as}:
\begin{equation}\label{eq:utilde}
	\tilde{\bu} = \bG^{c}(s) \tilde{\by} = \stackrel{l}{\underset{e=1}{\diag}}\{G^{c}_{e}(s)\}\tilde{\by},
\end{equation}
where each $G^{c}_{e}(s), e \in \{1,2,\dots,l\}$ is designed {\color{red}as an SNI controller}. The overall controller $\bG^{c}(s)$ is SNI {\color{red}which follows via a similar proof as the proof of} Lemma~\ref{lemma:overall_SNI}. 

\begin{remark}
	As shown in Fig.~\ref{fig:virtual}, each element $\tilde{\bu}_{e}$ in the input $\tilde{\bu}$ serves as a virtual transmission line in blue between neighboring buses $i$ and $j$. The overall input $\tilde{\bu}$ contributes to the stability of the overall feedback system (see in Section~\ref{sec:internal_stability}). {\color{red}Furthermore, if suitable angle measurements and battery storage elements are available, it would  be possible to add a virtual transmission line in the network (for a short period) where no physical transmission line existed.} {\color{green} Such virtual transmission lines can play a critical role in preventing events such as the South Australian blackout \cite{AEMO_SA} and in enabling transient stability to be maintained until orderly load shedding is carried out. This means that the proposed scheme has the potential to improve the resilience of the grid against natural disasters \cite{mohamed2019proactive} and even distributed cyber attacks \cite{chatterjee2017review}.}
\end{remark}
\medskip

\begin{figure}[tb]
	\centering
	\includegraphics[width=0.34 \linewidth]{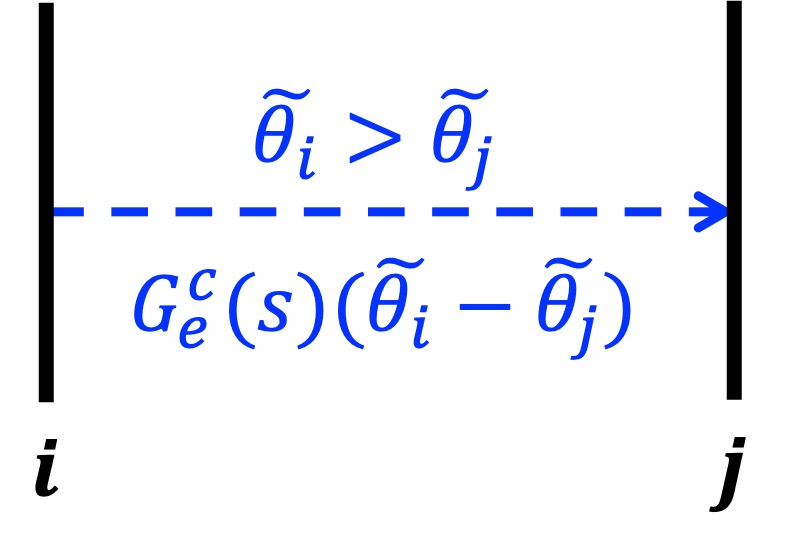}
	\caption{With edge $e \in \mcalE$ representing the true transmission line $(i,j)$, the input $\tilde{\bu}_{e}$ functions as a virtual transmission line between neighboring buses $i$ and $j$.}
	\label{fig:virtual}
\end{figure}

Finally, we synthesize $\tilde{\bu}$ into the real input $\bu \in \mathbb{R}^{n}$ that will be implemented in practice on each generator bus by using the local storage device:
\begin{equation}\label{eq:realu}
	\bu = \bQ \tilde{\bu}.
\end{equation}
By expanding~Eq.~\eqref{eq:realu}, the input for each generator bus $i \in \mcalV$ in a distributed manner is given by
\begin{equation}\label{eq:distributed_u}
	\bu_{i} =\sum_{j = 1}^{n} T_{ij}G^{c}_{e}(s)(\by_{i} - \by_{j}),
\end{equation}
where $e \in \{1,2,\dots,l\}$ represents the transmission line connecting neighboring buses $i$ and $j$.

\begin{remark}
	From~\eqref{eq:distributed_u}, we see that each generator bus only uses local information from its neighbors, i.e., the output $\by_{j \in \mcalN(i)}$, to produce its real input $\bu_{i}$.
\end{remark}
\medskip

In summary, the overall plant for the proposed feedback system is illustrated in Fig.~\ref{fig:overall_fb_plant}:
\begin{subequations}\label{eq:fb_sys}
	\begin{align}
		\dot{\bx}& = \bA\bx + \bB\bQ\tilde{\bu},\label{eq:post_x}\\
		\tilde{\by}& = \bQ^{T}\bC\bx,\label{eq:post_y}\\
		\tilde{\bu} &= \bG^{c}(s) \tilde{\by}.\label{eq:post_c}
	\end{align}
\end{subequations}

\begin{figure}[tb]
	\centering
	\includegraphics[width=0.45 \textwidth]{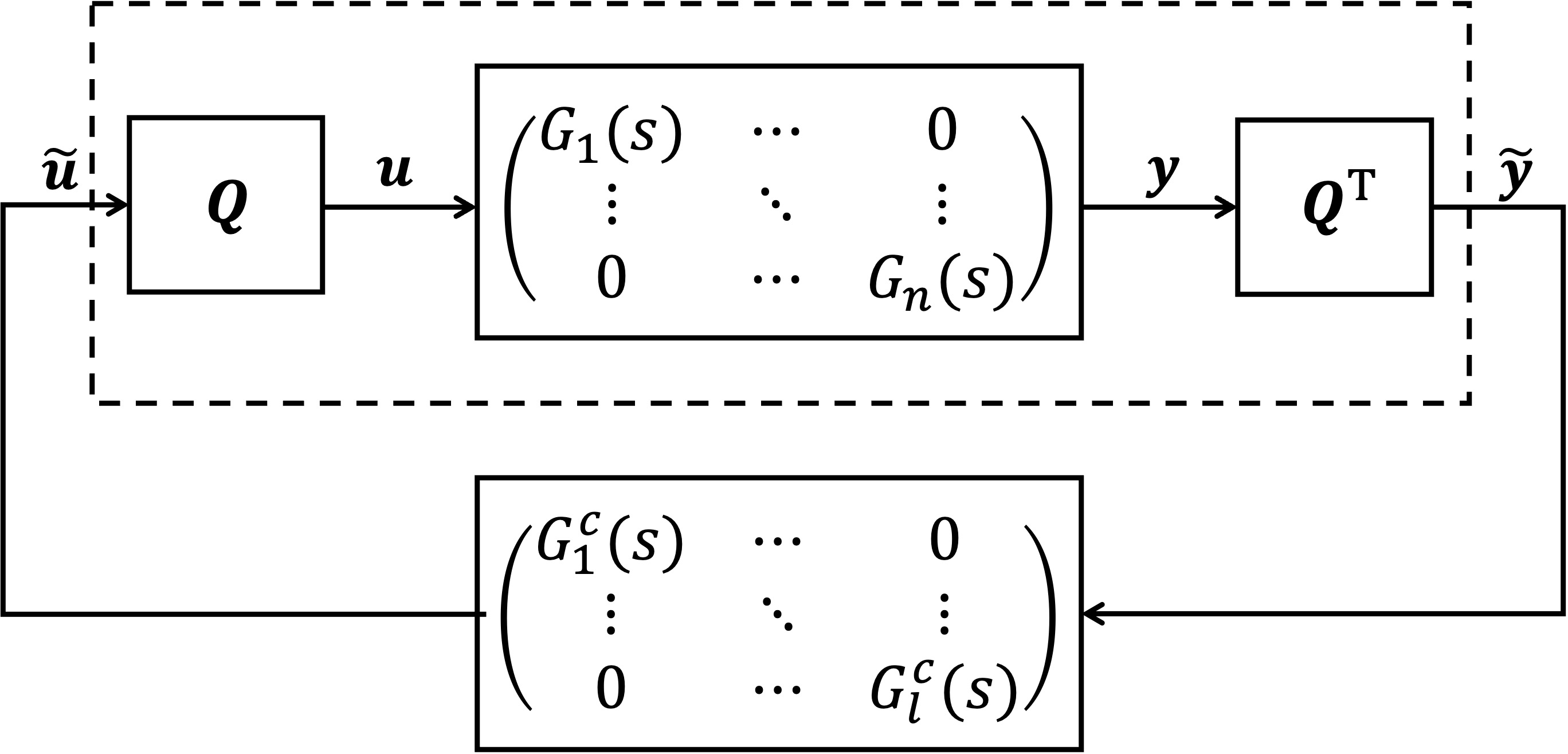}
	\vspace{0.5em}
	\caption{The overall plant for the proposed feedback system.}
	\vspace{-1em}
	\label{fig:overall_fb_plant}
\end{figure}

\subsection{Internal Stability}\label{sec:internal_stability}
In what follows, we show the internal stability of the positive interconnection of $\bQ^{T}\bG(s)\bQ$ and $\bG^{c}(s)$. The proof is inspired by \cite{wang2015robust}.
\medskip

The following lemma shows that the system~\eqref{eq:post_x} and \eqref{eq:post_y} as depicted in Fig. \ref{fig:overall_fb_plant} is an NI system.

\begin{lemma}
	The transfer function $\bQ^{T}\bG(s)\bQ$ is NI.
\end{lemma}
{\it Proof.} It is straightforward {\color{red}to show} that all the poles of $\bQ^{\top} \bG(s)\bQ$ lie in the OLHP. We then want to show that $\bj[\bQ^{\top}\bG (\bj \alpha) \bQ- \bQ^{\top} \bG^{\ast}(\bj \alpha)\bQ] \geq 0, \forall \alpha \geq 0$. Consider an arbitrary vector $\mathbf{v} \in \mathbb{R}^{l} $ and $\mathbf{v'} = \bQ\mathbf{v} \in \mathbb{R}^{n}$. It {\color{red}follows} that $\forall \alpha \geq 0$, 
\begin{align*}
	&\mathbf{v}^{\top}\bj[\bQ^{\top}\bG (\bj \alpha) \bQ- \bQ^{\top} \bG^{\ast}(\bj \alpha)\bQ] \mathbf{v} \\
	=\ &\mathbf{v}^{\top}\bQ^{\top} \bj[\bG (\bj \alpha) - \bG^{\ast}(\bj \alpha)]\bQ\mathbf{v}\\
	= \ & \mathbf{v'}^{\top} \bj[\bG (\bj \alpha) - \bG^{\ast}(\bj \alpha)]\mathbf{v'} \\
	\geq \ &  0.
\end{align*}
The last inequality comes from the fact that $\bG(s)$ is an SNI. The proof is now completed. \hfill$\square$

\medskip

\begin{theorem}\label{thm:internal_stability}
	Given the NI transfer function $\bQ^{T}\bG(s)\bQ$ of the post-processed system~\eqref{eq:post_x} and \eqref{eq:post_y} and the SNI transfer function $\bG^{c}(s)$ of the feedback controller~\eqref{eq:post_c} with $\bQ^{T}\bG(\infty)\bQ\bG^{c}(\infty) = 0$ and $\bG^{c}(\infty) \geq 0 $, the positive-feedback interconnection~\eqref{eq:fb_sys} of $\bQ^{T}\bG(s)\bQ$ and $\bG^{c}(s)$ is internally stable if and only if 
	\begin{equation}\label{eq:internal_stability_condition}
		\lambda_{\max}(\bQ^{T}\bG(0)\bQ\bG^{c}(0)) < 1.
	\end{equation}
\end{theorem}

{\it \noindent Proof.} Theorem~\ref{thm:internal_stability} {\color{red}follows directly from} Lemma~\ref{lemma:SNI}.\hfill$\square$

\medskip

\begin{proposition}\label{prop:internal_stability}
	Given the NI transfer function $\bQ^{T}\bG(s)\bQ$ of the post-processed system~\eqref{eq:post_x} and \eqref{eq:post_y} and the SNI transfer function $\bG^{c}(s)$ of the feedback controller~\eqref{eq:post_c} with $\bQ^{T}\bG(\infty)\bQ\bG^{c}(\infty) = 0$ and $\bG^{c}(\infty) \geq 0 $, the positive-feedback interconnection~\eqref{eq:fb_sys} of $\bQ^{T}\bG(s)\bQ$ and $\bG^{c}(s)$ is internally stable if
	\begin{equation}\label{eq:prop}
		\lambda_{\max}(\bG(0))\lambda_{\max}(\bG^{c}(0)) < \frac{1}{\lambda_{\max}(\bQ\bQ^{T})}.
	\end{equation}
\end{proposition}

{\it \noindent Proof.} We begin by making some {\color{red}observations}. First, we have $\bG^{c}(0) \geq \bG^{c}(\infty) \geq 0$, which comes from Lemma~\ref{lemma:zero_frequency} and the given condition $\bG^{c}(\infty) \geq 0$. Second, since $\bQ^{T}\bG(0)\bQ \geq 0$  and $\bG^{c}(0) \geq 0$, Lemma~\ref{lemma:lambda_inequality} implies that $\lambda_{\max}(\bQ^{T}\bG(0)\bQ\bG^{c}(0)) \leq \lambda_{\max}(\bQ^{T}\bG(0)\bQ) \lambda_{\max}(\bG^{c}(0)) $. Third, we have $\lambda_{\max}(\bQ^{\top}\bG(0)\bQ) \leq \lambda_{\max}(\bG(0)) \lambda_{\max}(\bQ^{\top}\bQ) $ due to the fact that $\bQ^{\top}(\bG(0) - \lambda_{\max}(\bG(0))\mathbf{I})\bQ \leq 0$. Therefore, {\color{red}using} Eq.~\eqref{eq:prop}, we obtain that 
\begin{align*}
	&\lambda_{\max}(\bQ^{T}\bG(0)\bQ\bG^{c}(0)) \\
	\leq \ & \lambda_{\max}(\bQ^{T}\bG(0)\bQ) \lambda_{\max}(\bG^{c}(0))  \\
	\leq \ & \lambda_{\max}(\bG(0))\lambda_{\max}(\bQ^{T}\bQ)\lambda_{\max}(\bG^{c}(0))\\
	< \ & 1.
\end{align*}
Finally, Lemma~\ref{thm:internal_stability} {\color{red}is applied to conclude internal stability}. The proof is now completed. \hfill$\square$

\medskip

\begin{remark}
	Theorem~\ref{thm:internal_stability} and Proposition~\ref{prop:internal_stability} imply that as long as an SNI controller is applied to the transmission network, the state of the overall feedback system~\eqref{eq:fb_sys} will eventually reach a stable equilibrium at $\mathbf{0}_{2n}$. It further {\color{red}follows that} $\omega_{i} \to \omega^{0}, \forall i \in \mcalV$ from Eq.~\eqref{eq:frequency_def} and $\theta_{i}- \theta_{j} \to \bar{\psi}_{ij}, \forall (i,j) \in \mcalE$ from Eqs.~\eqref{eq:consensus} and \eqref{eq:nominal_diff}, as time goes to infinity.
\end{remark}
\medskip

In summary, we  {\color{red}propose angle based} feedback controllers {\color{red}using} local storage devices on generator buses as actuators. In particular, the proposed controllers have three advantages: 1) based on  negative-imaginary systems theory, the interconnection of the power transmission network and the {\color{red}angle based} feedback controllers is proved to be internally stable; 2) the frequencies {\color{red}of the} different generator buses are synchronized to a nominal value and the power flows on transmission lines are maintained at {\color{red}steady-state} levels as observed before a fault; 3) with local measurement and local communication of rotor angle data, {\color{red}the overall control system} operates in a fully distributed manner. {\color{green}Note that the reliable operation of such a grid control system will require extremely reliable angle measurements and communications. This could be achieved by using aerospace industry ideas such as  triple redundancy in angle measurements and communications. Thus, we are effectively proposing to ``fly the grid by wire''.}

\begin{figure}[tb]
	\centering
	\includegraphics[width=0.4\linewidth]{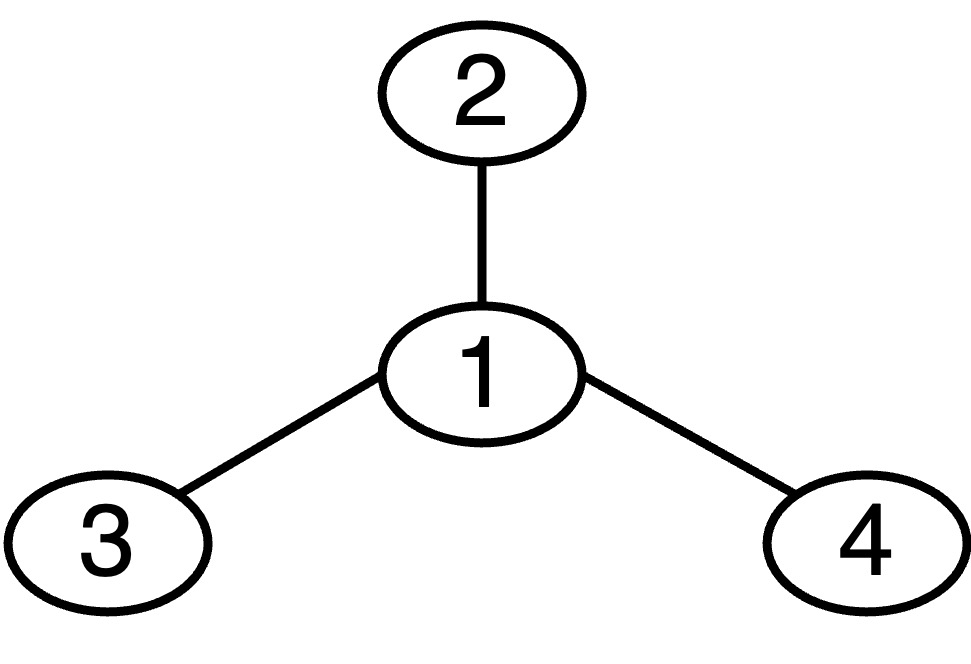}
	\caption{A 4-generator-bus transmission network.}
	\label{fig:4-generator-bus}
\end{figure}

\begin{figure*}[tb]
	\centering
	\subfloat[The transmission line $(1,2)$]{\includegraphics[width=.33\linewidth]{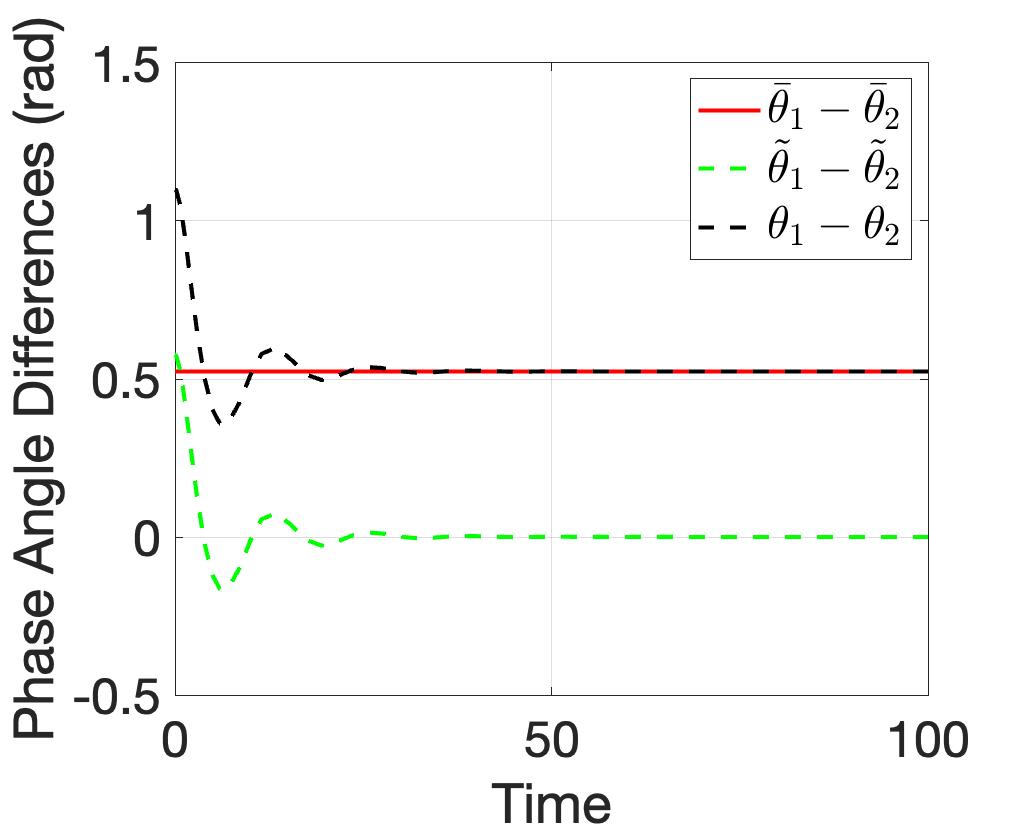}}
	\subfloat[The transmission line $(1,3)$]{\includegraphics[width=.33\linewidth]{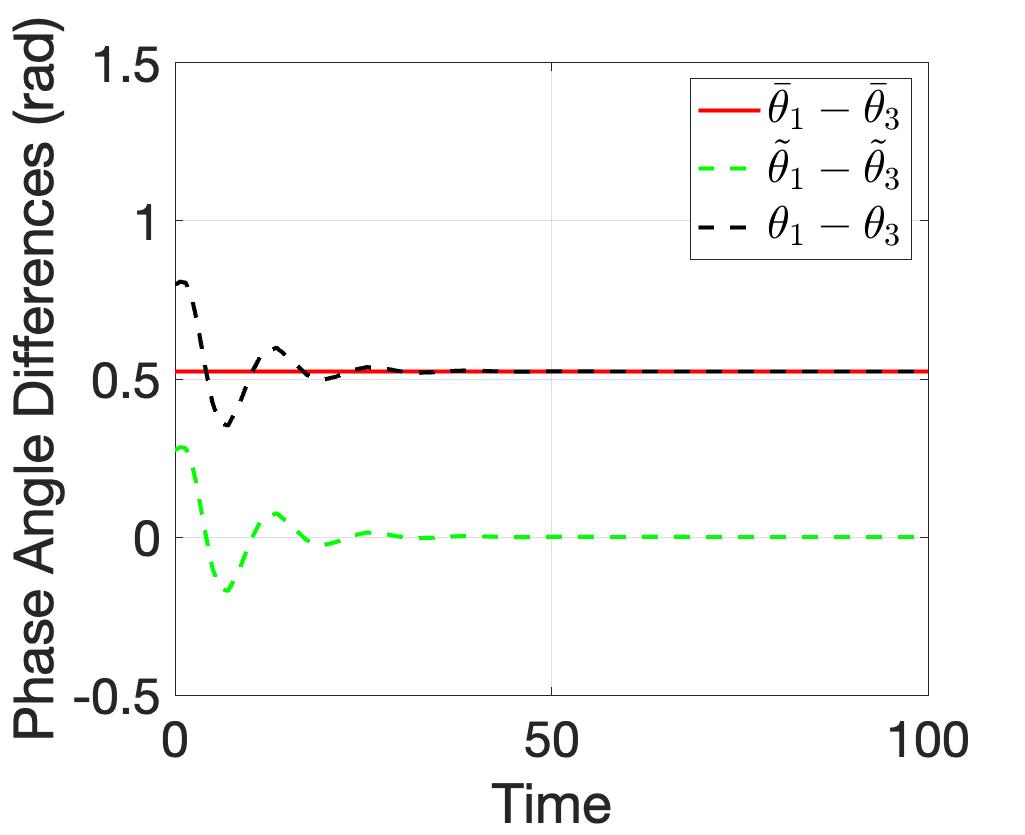}} 
	\subfloat[The transmission line $(1,4)$]{\includegraphics[width=.33\linewidth]{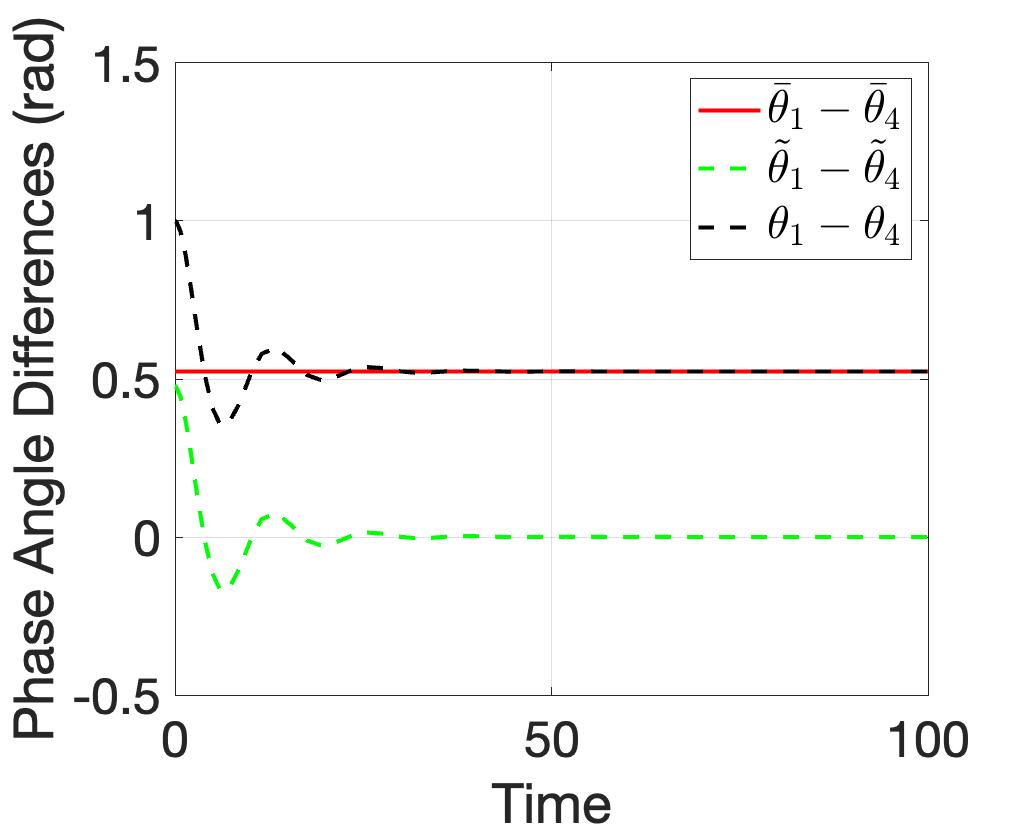}}
	\caption{The comparison of $\bar{\delta}_{i}-\bar{\delta}_{j}$, $\tilde{\delta}_{i}-\tilde{\delta}_{j}$, and $\delta_{i}-\delta_{j}$ for all $(i,j) \in \mcalE$ in the 4-generator-bus transmission network.}
	\label{fig:comparison}
\end{figure*}

\begin{figure}[tb]
	\centering
	\includegraphics[width=0.4 \textwidth]{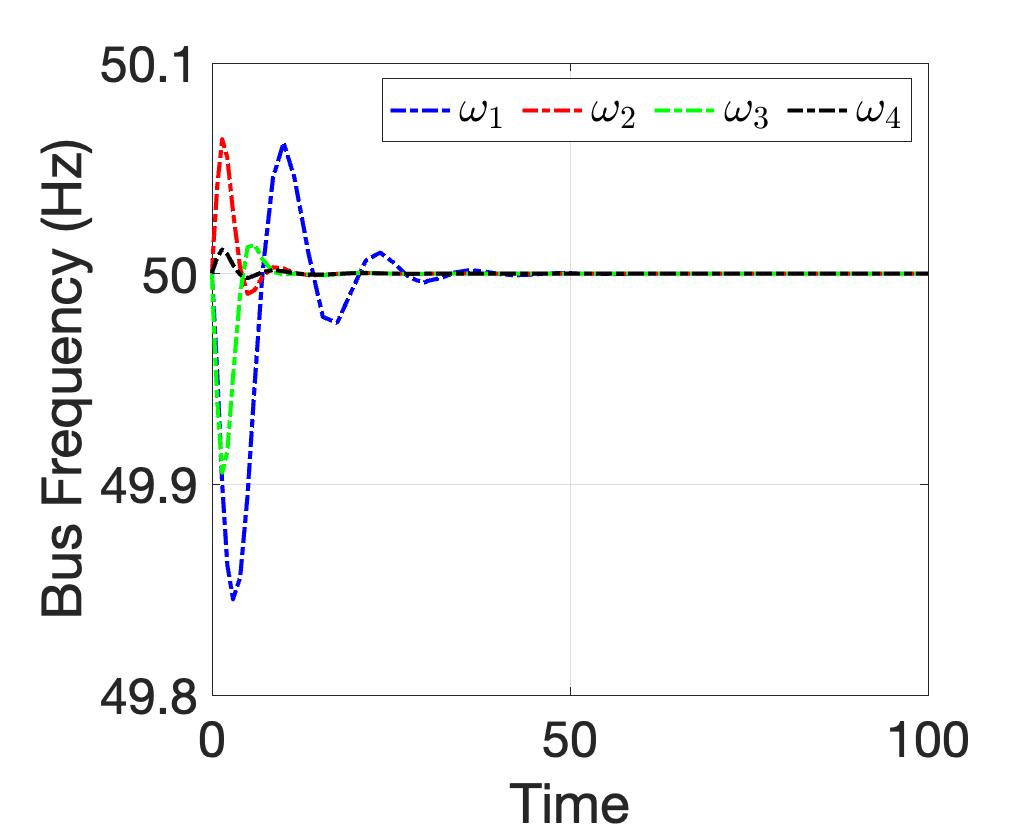}
	\caption{The rotor angle {\color{red}frequencies of the} four buses.}
	\label{fig:frequency_synchronization}
\end{figure}

\section{Illustrative Examples}\label{sec:examples}
Consider a 4-generator-bus transmission network with a star topology as illustrated in Fig.~\ref{fig:4-generator-bus}. The maximum eigenvalue of $\bQ\bQ^{\top}$ is computed as $4$. The system  frequency and angles at steady state before the occurrence of a fault are as follows: 
\begin{align*}
	&\omega^{0} = 50 \ \mathrm{Hz},\\
	&\bar{\theta}_{i \in\mcalV}= \left(\frac{\pi}{3}, \frac{\pi}{6}, \frac{\pi}{6},\frac{\pi}{6}\right).
\end{align*}
We then consider the occurrence of a fault with randomly generated phase and power flow deviations.
\medskip

{\noindent \bf Generator and Controller Settings.} The parameters of generator buses are set as 
$\{M_{i}\}_{i \in \mcalV} = (20, 5, 5, 5)$,
$\{D_{i}\}_{i \in \mcalV}=(0.2,0.2,0.2,0.2)$, and
$\{K_{i}\}_{i \in \mcalV} = (1,1,1,1)$.
Thus, the transfer function is described by  $\bG^{s} = \diag\{\frac{0.05}{s^2 + 0.25s + 0.25},\frac{0.2}{s^2 + s + 1},\frac{0.2}{s^2 + s + 1},\frac{0.2}{s^2 + s + 1}\}$.
We select an SNI controller $\diag\{\frac{0.4}{10s+1}, \frac{0.5}{10s+1},\frac{0.3}{10s+1}\}$.
We can check that $\bQ^{T}\bG(\infty)\bQ\bG^{c}(\infty) = 0$ and $\bG^{c}(\infty) \geq 0 $. We can also check that $ \lambda_{\max}(\bG(0))\lambda_{\max}(\bG^{c}(0)) = 0.2 \times 0.5 = 0.1 < \frac{1}{4} =  \frac{1}{\lambda_{\max}(\bQ\bQ^{T})}$.
\medskip

{\bf \noindent Simulation Results.} In Fig.~\ref{fig:frequency_synchronization}, we plot rotor angle {\color{red}frequencies of the} four buses. The bus frequencies are regulated at the nominal level of $50$ Hz. In Fig.~\ref{fig:comparison}, we {\color{red}present} the comparison of $\bar{\theta}_{i}-\bar{\theta}_{j}$, $\tilde{\theta}_{i}-\tilde{\theta}_{j}$, and $\theta_{i}-\theta_{j}$ for all $(i,j) \in \mcalE$ in the underlying 4-generator-bus transmission network.  {\color{red}It is shown that} $\tilde{\theta_{1}} - \tilde{\theta_{j}}, j= 2,3,4$ reach consensus at $0$. The trajectories of $\theta_{1}-\theta_{j}, j = 2,3,4,$ over time approach the flat line $\bar{\theta}_{1}-\bar{\theta}_{j}, j = 2,3,4$, which implies that the {\color{red}steady-state} angle differences between neighboring buses in the transmission network are maintained {\color{red}using} our design. Both results validate Theorem~\ref{thm:internal_stability} and Proposition~\ref{prop:internal_stability} and show the merits of our proposed SNI controller.

\section{Conclusion}\label{sec:conclusion}
This paper considered a power transmission network described by interconnected nonlinear swing dynamics on generator buses. We assumed {\color{red}that real time measurements} of the rotor angle for each generator bus {\color{red}are available} and {\color{red}that all} generator buses {\color{red}have large-scale} energy storage batteries {\color{red}that serve as controller actuators}. Based on negative-imaginary systems theory, the {\color{red}angle based} feedback controllers were designed and  implemented in a distributed manner with local information. Our analysis demonstrated the internal stability of the interconnection between the power transmission network and the {\color{red}angle based} feedback controllers, which implied the advantages of achieving frequency synchronization and preserving {\color{red}steady-state} power flows on transmission lines. {\color{red}Also, the controllers augmented the transient stability of the system.} Simulation results confirmed the validity of our analysis. Taking into account the saturation of local storage device {\color{red}actuators} is a {\color{red}possible} future {\color{red}research} direction. Future research will also investigate how the current grid can be transitioned into the proposed system, one transmission line at a time, as they reach their capacity.

	\bibliographystyle{IEEEtran}
	\bibliography{ref}

\begin{thebibliography}{10}
\providecommand{\url}[1]{#1}
\csname url@samestyle\endcsname
\providecommand{\newblock}{\relax}
\providecommand{\bibinfo}[2]{#2}
\providecommand{\BIBentrySTDinterwordspacing}{\spaceskip=0pt\relax}
\providecommand{\BIBentryALTinterwordstretchfactor}{4}
\providecommand{\BIBentryALTinterwordspacing}{\spaceskip=\fontdimen2\font plus
\BIBentryALTinterwordstretchfactor\fontdimen3\font minus
  \fontdimen4\font\relax}
\providecommand{\BIBforeignlanguage}[2]{{%
\expandafter\ifx\csname l@#1\endcsname\relax
\typeout{** WARNING: IEEEtran.bst: No hyphenation pattern has been}%
\typeout{** loaded for the language `#1'. Using the pattern for}%
\typeout{** the default language instead.}%
\else
\language=\csname l@#1\endcsname
\fi
#2}}
\providecommand{\BIBdecl}{\relax}
\BIBdecl

\bibitem{AEMO_TE}
\BIBentryALTinterwordspacing
{Australian Energy Market Operator (AEMO)}. {2023 Transmission Expansion
  Options Report}. [Online]. Available: \url{https://www.aemo.com.au}
\BIBentrySTDinterwordspacing

\bibitem{petersen2010feedback}
I.~R. Petersen and A.~Lanzon, ``Feedback control of negative-imaginary
  systems,'' \emph{IEEE Control Systems Magazine}, vol.~30, no.~5, pp. 54--72,
  2010.

\bibitem{lanzon2008stability}
A.~Lanzon and I.~R. Petersen, ``Stability robustness of a feedback
  interconnection of systems with negative imaginary frequency response,''
  \emph{IEEE Transactions on Automatic Control}, vol.~53, no.~4, pp.
  1042--1046, 2008.

\bibitem{xiong2010negative}
J.~Xiong, I.~R. Petersen, and A.~Lanzon, ``A negative imaginary lemma and the
  stability of interconnections of linear negative imaginary systems,''
  \emph{IEEE Transactions on Automatic Control}, vol.~55, no.~10, pp.
  2342--2347, 2010.

\bibitem{petersen2016negative}
I.~R. Petersen, ``Negative imaginary systems theory and applications,''
  \emph{Annual Reviews in Control}, vol.~42, pp. 309--318, 2016.

\bibitem{wang2015robust}
J.~Wang, A.~Lanzon, and I.~R. Petersen, ``Robust cooperative control of
  multiple heterogeneous negative-imaginary systems,'' \emph{Automatica},
  vol.~61, pp. 64--72, 2015.

\bibitem{bevrani2014robust}
H.~Bevrani, \emph{Robust power system frequency control}.\hskip 1em plus 0.5em
  minus 0.4em\relax Springer, 2014, vol.~4.

\bibitem{ilic2012toward}
M.~D. Ili{\'c} and Q.~Liu, ``Toward sensing, communications and control
  architectures for frequency regulation in systems with highly variable
  resources,'' \emph{Control and optimization methods for electric smart
  grids}, pp. 3--33, 2012.

\bibitem{wood2013power}
A.~J. Wood, B.~F. Wollenberg, and G.~B. Shebl{\'e}, \emph{Power generation,
  operation, and control}.\hskip 1em plus 0.5em minus 0.4em\relax John Wiley \&
  Sons, 2013.

\bibitem{liu2013decentralized}
H.~Liu, Z.~Hu, Y.~Song, and J.~Lin, ``Decentralized vehicle-to-grid control for
  primary frequency regulation considering charging demands,'' \emph{IEEE
  Transactions on Power Systems}, vol.~28, no.~3, pp. 3480--3489, 2013.

\bibitem{beil2016frequency}
I.~Beil, I.~Hiskens, and S.~Backhaus, ``Frequency regulation from commercial
  building {HVAC} demand response,'' \emph{Proceedings of the IEEE}, vol. 104,
  no.~4, pp. 745--757, 2016.

\bibitem{kim2014modeling}
Y.-J. Kim, L.~K. Norford, and J.~L. Kirtley, ``Modeling and analysis of a
  variable speed heat pump for frequency regulation through direct load
  control,'' \emph{IEEE Transactions on Power Systems}, vol.~30, no.~1, pp.
  397--408, 2014.

\bibitem{gonen2009electrical}
T.~Gonen, \emph{Electrical power transmission system engineering: analysis and
  design}.\hskip 1em plus 0.5em minus 0.4em\relax CRC Press, 2009.

\bibitem{doukas2011electric}
H.~Doukas, C.~Karakosta, A.~Flamos, and J.~Psarras, ``Electric power
  transmission: An overview of associated burdens,'' \emph{International
  journal of energy research}, vol.~35, no.~11, pp. 979--988, 2011.

\bibitem{tran2019efficient}
V.~T. Tran, M.~R. Islam, K.~M. Muttaqi, and D.~Sutanto, ``An efficient energy
  management approach for a solar-powered {EV} battery charging facility to
  support distribution grids,'' \emph{IEEE Transactions on Industry
  Applications}, vol.~55, no.~6, pp. 6517--6526, 2019.

\bibitem{borenstein2022s}
S.~Borenstein, ``It’s time for rooftop solar to compete with other
  renewables,'' \emph{Nature Energy}, vol.~7, no.~4, pp. 298--298, 2022.

\bibitem{liu2015comparison}
J.~Liu, Y.~Miura, and T.~Ise, ``Comparison of dynamic characteristics between
  virtual synchronous generator and droop control in inverter-based distributed
  generators,'' \emph{IEEE Transactions on Power Electronics}, vol.~31, no.~5,
  pp. 3600--3611, 2015.

\bibitem{vivsic2020synchronous}
I.~Vi{\v{s}}i{\'c}, I.~Strnad, and A.~Maru{\v{s}}i{\'c}, ``Synchronous
  generator out of step detection using real time load angle data,''
  \emph{Energies}, vol.~13, no.~13, p. 3336, 2020.

\bibitem{kundur2022power}
P.~S. Kundur and O.~P. Malik, \emph{Power system stability and control}.\hskip
  1em plus 0.5em minus 0.4em\relax McGraw-Hill Education, 2022.

\bibitem{zhao2012swing}
C.~Zhao, U.~Topcu, and S.~Low, ``Swing dynamics as primal-dual algorithm for
  optimal load control,'' in \emph{IEEE Third International Conference on Smart
  Grid Communications}, 2012, pp. 570--575.

\bibitem{rathnayake2022multivariable}
D.~B. Rathnayake and B.~Bahrani, ``Multivariable control design for
  grid-forming inverters with decoupled active and reactive power loops,''
  \emph{IEEE Transactions on Power Electronics}, vol.~38, no.~2, pp.
  1635--1649, 2022.

\bibitem{lasseter2019grid}
R.~H. Lasseter, Z.~Chen, and D.~Pattabiraman, ``Grid-forming inverters: A
  critical asset for the power grid,'' \emph{IEEE Journal of Emerging and
  Selected Topics in Power Electronics}, vol.~8, no.~2, pp. 925--935, 2019.

\bibitem{chen2020distributed}
X.~Chen, C.~Zhao, and N.~Li, ``Distributed automatic load frequency control
  with optimality in power systems,'' \emph{IEEE Transactions on Control of
  Network Systems}, vol.~8, no.~1, pp. 307--318, 2020.

\bibitem{dorfler2012synchronization}
F.~Dorfler and F.~Bullo, ``Synchronization and transient stability in power
  networks and nonuniform {Kuramoto} oscillators,'' \emph{SIAM Journal on
  Control and Optimization}, vol.~50, no.~3, pp. 1616--1642, 2012.

\bibitem{machowski1997power}
J.~Machowski, J.~W. Bialek, and J.~R. Bumby, \emph{Power system dynamics and
  stability}.\hskip 1em plus 0.5em minus 0.4em\relax John Wiley \& Sons, 1997.

\bibitem{AEMO_SA}
\BIBentryALTinterwordspacing
{Australian Energy Market Operator (AEMO)}. {Potential Loss of Multiple
  Generators in South Australia}. [Online]. Available:
  \url{https://www.aemo.com.au}
\BIBentrySTDinterwordspacing

\bibitem{mohamed2019proactive}
M.~A. Mohamed, T.~Chen, W.~Su, and T.~Jin, ``Proactive resilience of power
  systems against natural disasters: A literature review,'' \emph{IEEE Access},
  vol.~7, pp. 163\,778--163\,795, 2019.

\bibitem{chatterjee2017review}
K.~Chatterjee, V.~Padmini, and S.~Khaparde, ``Review of cyber attacks on power
  system operations,'' in \emph{IEEE Region 10 Symposium}, 2017, pp. 1--6.

\end{thebibliography}

\end{document}